# Fraunhofer patterns in magnetic Josephson junctions with non-uniform magnetic susceptibility

B. Börcsök[1], S. Komori[1], A. I. Buzdin[1, 2], and J. W. A. Robinson[1, ]*

*To whom correspondence should be addressed: jjr33@cam.ac.uk

[1] *Department of Materials Science & Metallurgy, University of Cambridge, 27 Charles Babbage Road, Cambridge CB3 0FS, United Kingdom*

[2] *Université Bordeaux, CNRS, LOMA, UMR- 5798, F-33400 Talence, France*

**Abstract**

The development of superconducting memory and logic based on magnetic Josephson junctions relies on an understanding of junction properties and, in particular, the dependence of critical current on external magnetic flux (i.e. Fraunhofer patterns). With the rapid development of Josephson junctions with various forms of inhomogeneous barrier magnetism, Fraunhofer patterns are increasingly complex. In this paper we model Fraunhofer patterns for magnetic Josephson junctions in which the barrier magnetic susceptibility is position- and external magnetic field dependent. The model predicts anomalous Fraunhofer patterns in which local minima in the Josephson critical current can be nonzero and non-periodic with external magnetic flux due to an interference effect between highly magnetised and demagnetised regions.

**Introduction**

S-wave singlet superconductivity and ferromagnetism are competing phases. Over the past half century considerable research has been undertaken in order to understand the interaction between these phenomena at superconductor/ferromagnet (S/F) interfaces[1–7]. A key experimental development was the demonstration of F-thickness-dependent oscillations in the Josephson critical current $I_c$ in S/F/S junctions, first using weak ferromagnets (CuNi and PdNi[8–13]) and then strong ferromagnets (Fe, Co, Ni and NiFe;[14–19]). This behaviour is a manifestation of the magnetic exchange field acting differentially on the spins of the singlet pairs, which induces oscillations in the superconducting order parameter in F superimposed on a rapid decay with a singlet coherence length of $\xi_s$ < 3 nm[10,15,17,19]. The superconductivity in F can also be detected via tunnelling density of states (DoS) measurements[20,21] and point contract Andreev spectroscopy[22,23]. Furthermore, the magnetic exchange fields from F induces a Zeeman-like splitting of the DoS in S close to the S/F interface[24–26].

Recently there is a focus on Josephson junctions with inhomogeneous barrier magnetism, involving misaligned F layers[27–32] and/or rare earth magnets such as Ho or Gd[33,34], in order to transform singlet pairs into spin-aligned triplet pairs[3,6,35]. Triplet pairs are spin-polarized and stable in a magnetic exchange field and decay in Fs over length scales exceeding $\xi_s$[3,5]. However, the relatively large (total) magnetic barrier thickness in triplet junctions introduces significant flux which, in combination with magnetic inhomogeneity, creates a complex dependence of $I_c$ on external magnetic field $H$[36,37].

A complication for junctions with magnetically inhomogeneous rare earths such as Ho (or Er) relates to the fact that the magnetic ordering and local magnetic susceptibility $\chi$ depends on a competition between Ruderman–Kittel–Kasuya–Yosida (RKKY) coupling between localized moments and shape anisotropy[38]. Let us take Ho an example. In single crystals the moments align into an antiferromagnetic spiral below 133 K made up of F-ordered basal planes with moments in successive planes rotated 30° relative to each other due to the RKKY coupling[39,40]. Below about 20 K the moments in Ho tilt slight out-of-plane although this is not observed in thin film due to strain[41]. The antiferromagnetic spiral has a zero net magnetic moment but applying magnetic fields parallel to the basal planes[42,43] induces an irreversible transition to a ferromagnetic state. In epitaxial thin-films, similar properties are reproduced although the antiferromagnetic spiral can remain stable over a wide field range[44]. In textured or polycrystalline thin films the antiferromagnetic spiral can even remain fully reversible even after applying magnetically saturating fields[45]. At the edges of Ho, however, RKKY coupling is reduced which, in combination with shape anisotropy, may favour easy magnetization alignment along edge regions. This translates to localized enhancements in $\chi$ at edges and thus an inhomogeneous magnetic induction in the junction.

In this paper we calculate the magnetic-field-dependence of the maximum Josephson critical current $I_c$ (i.e. Fraunhofer patterns) in S/F/S junctions with a position- and magnetic-field-dependent-$\chi$ (Fig. 1). The model predicts anomalous Fraunhofer patterns due to spatial variations in $\chi$ and magnetic induction in which local minima in $I_c(H)$ can be nonzero and non-periodic due to interface between highly magnetised and demagnetised regions.

The S/F/S junction geometry under consideration is sketched in Fig. 1 which summarizes the magnetization process [Fig. 1(a)-1(d)]. We consider the case of a Josephson junction with a width $L$ that is smaller than the Josephson penetration depth (which is usually the case for experiments), so the applied magnetic field $H$ fully penetrates the barrier[46]. Following standard procedures (see e.g. [47]) we calculate the phase variation across the S/F/S barrier taking into account the contribution from the magnetic moment to the total flux through the junction during the magnetization process as summarised in Fig. 1(e). Applying $H$ parallel to y causes the magnetization $M$ along junction edges parallel to y to propagate inwards towards the junction centre until magnetic saturation $H = H_s$. The expansion of the magnetized region is assumed to be reversible with a width that depends on $H$ and not magnetic field history. The propagation rate of the magnetized region is linear with $H$ in our model and the position of the boundary between magnetized and demagnetized regions is $a = L/2 - PH$ (where $P$ is the propagation parameter and $L$ the junction width). The magnetization is uniform in the y direction and position-dependent in the x direction with $M(x) = \chi(x)H$. We note that for certain materials the propagation rate of the magnetized region with $H$ may not be linear, but as a first approximation which choose a linear form here which is likely to apply more broadly to magnetic materials in nanopillar Josephson junctions.

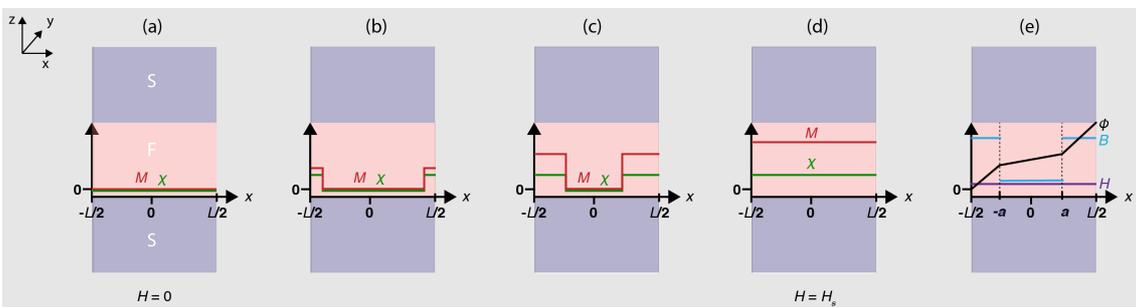

**Fig. 1. Magnetization process for an S/F/S Josephson junction with a position ($x$) and magnetic field ($H$) dependent magnetic susceptibility $\chi(x, H)$ and magnetization $M(x, H)$.** (a) For $H = 0$ the net barrier moment is zero everywhere but on increasing $H$ (b-d), $M$ increases faster at the junction edges and propagates inwards until the magnetic barrier saturates ($H = H_s$) (d). (e) Spatial variation of magnetic induction $B$ and superconducting phase difference $\phi$ for $0 < H < H_s$. The external field $H$ is always applied in the y direction. The variables ($M$, $\phi$, $H$, $B$) plotted on the

z-axis are labelled.

The spatial variation *M(x)* means that the magnetic induction *B(x)* is non-uniform. The line integral of *B(x)* across the junction gives the spatial gradient of the superconducting phase

$$\frac{\partial \varphi}{\partial x} = \frac{2\pi \int B dz}{\Phi_0} = \frac{2\pi[\bar{d}H + 4\pi\chi dH]}{\Phi_0}, \qquad (1)$$

where $\Phi_0$ is the flux quanta [$h/(2e) \approx 2.06 \times 10^{-15}$ Wb] and $\bar{d} = d + 2\lambda$ is the effective junction thickness. Hence, $\varphi(x)$ in the magnetized ($a < x < L/2$) and demagnetized ($|x| < a$) regions is given by

$$\varphi(x) = \frac{2\pi\Phi}{\Phi_0}\frac{x}{L} + \varphi_0, \qquad |x| < a \quad \text{where} \quad \Phi = HL\bar{d}, \qquad (2)$$

$$\varphi(x) = \frac{2\pi\Phi}{\Phi_0}\frac{(x-a)}{L}\left(1 + 4\pi\chi_0\frac{d}{\bar{d}}\right) + \frac{2\pi\Phi}{\Phi_0}\frac{a}{L} + \varphi_0, \quad a < x < L/2 \qquad (3)$$

where $\varphi_0$ is a constant that is set to give the maximum total critical current through the junction. The second term in equation (3) ensures $\varphi(x)$ is continuous. The spatial variation of the magnetic parameters and the superconducting phase difference are sketched in Fig. 1(e).

The position-dependent current density *j(x)* in the magnetized and demagnetized regions are

$$j(x) = j_c * \sin\left[\frac{2\pi\Phi}{\Phi_0}\frac{x}{L} + \varphi_0\right], \text{ for } |x| < a \qquad (4)$$

$$j(x) = j_c * Q * \sin\left[\frac{2\pi\Phi}{\Phi_0}\frac{(x-a)}{L}\left(1 + 4\pi\chi_0\frac{d}{\bar{d}}\right) + \frac{2\pi\Phi}{\Phi_0}\frac{a}{L} + \varphi_0\right], \text{ for } a < x < \frac{L}{2} \qquad (5)$$

where $j_c$ is the maximum critical current density in the demagnetized region and *Q* is the ratio of the critical current densities in the magnetized and demagnetized regions - i.e. $Q = j_{m,c}/j_c$. The net exchange field in the magnetised regions can favour a transition to a π-state[48,49] and hence the directions of $j_c$ and $j_{m,c}$ can be opposite to each other meaning can *Q* be negative. The total critical current through the junction is thus $I = \int_{-\frac{w}{2}}^{\frac{w}{2}} \int_{-L/2}^{L/2} j(x) dx \, dy = w \int_{-L/2}^{L/2} j(x) dx$, where *w* is the junction width in the *y* direction. From symmetry, the maximum critical current is therefore achieved by setting $\varphi_0 = \pi/2$ which yields

$$I_c(f) = 2w \int_0^{\frac{L}{2}} j(x) dx =$$

$$2wj_c \left\{ \int_0^a \sin\left[\frac{2\pi\Phi}{\Phi_0}\frac{x}{L} + \frac{\pi}{2}\right] dx + Q \int_a^{\frac{L}{2}} \sin\left[\frac{2\pi\Phi}{\Phi_0}\frac{(x-a)}{L}\left(1 + 4\pi\chi_0\frac{d}{\tilde{d}}\right) + \frac{2\pi\Phi}{\Phi_0}\frac{a}{L} + \frac{\pi}{2}\right] dx \right\}. \quad (6)$$

To illustrate the general features of our model, we introduce the following dimensionless parameters: the relative position of the boundary between the magnetised and demagnetised regions $l = a/L$, the effective permeability $q = 4\pi\chi_0\frac{d}{\tilde{d}}+1$, and the normalised flux $f = \frac{\Phi}{\Phi_0}$.

Substituting these parameters into equation (6) gives the following expression for $I_c$

$$I_c = 2I_{c0} \int_0^l \cos[2\pi f \tilde{x}] \, d\tilde{x} + 2I_{c0} Q \int_l^{\frac{1}{2}} \cos[2\pi f (\tilde{x} - l)q + 2\pi f l] \, d\tilde{x}, \quad (7)$$

where $I_{c0} = Lwj_c$ is the H = 0 total critical current of the junction and $l = 0.5 - pf$ with $p = \frac{\Phi_0}{\tilde{d}L^2}P$. Calculating the integrals analytically we obtain

$$I_c = \frac{I_{c0}}{\pi f} \left\{ \sin(2\pi f(1/2 - pf)) + \frac{Q}{q}\left[\sin(2\pi f(1/2 + pf(q-1))) - \sin(2\pi f(1/2 - pf))\right] \right\} \quad (8)$$

$$= \frac{I_{c0}}{\pi f} \left[ \sin(\pi f(1 - 2pf)) + 2\frac{Q}{q}\sin(\pi f^2 pq)) \cos(\pi f(1 + pf(q-2))) \right]. \quad (9)$$

For $p = 0$, meaning the junction is demagnetised for all values of H, we recover the standard Fraunhofer relation $I_c(f) = I_{c0}\frac{\sin \pi f}{\pi f}$. The solution takes the same form when the magnetised and the demagnetised regions are equivalent – i.e. $Q = 1$ and $q = 1$ for all values of p.

At magnetic saturation f is 1/2p meaning equation (7) is only valid for $|f|<1/2p$. For $|f| \geq 1/2p$, the barrier is magnetised with a high effective permeability q with $I_c(f) = I_{cm}\frac{\sin(\pi f q)}{(\pi f q)}$, where $I_{c\,m}$ is the total critical current in the magnetized state and $I_{c\,m} = Lwj_{c\,m} = LwQj_c$. The shape of $I_c(f)$ is thus determined by $Q$, $p$ and $q$ and its magnitude by $j_c$ and the junction area. In Fig. 2 we have plotted example $I_c(f)$ patterns.

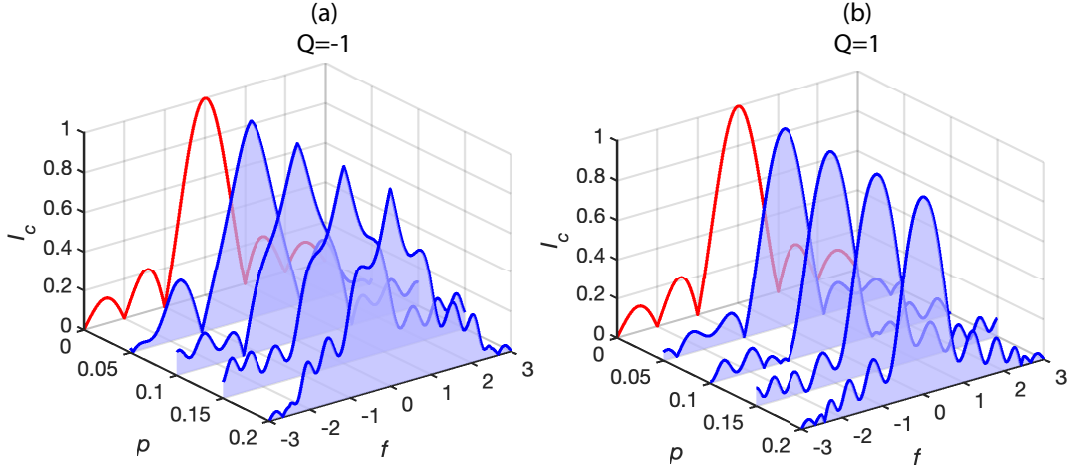

**Fig. 2. $I_c(f)$ vs $p$ and $f$ for positive and negative Q.** The blue curves show $I_c(f)$ for $q = 3, Q = -1$ (a) and $Q = 1$ (b). The red curves show standard Fraunhofer patterns for a demagnetized junction ($p$ = 0).

When the susceptibilities in the magnetized and demagnetized regions are different, we observe an interference in the critical current. However, due to the movement of the boundary between the magnetized and demagnetized regions with field, $I_c(f)$ is more complicated than simply the superposition of two sinc functions. Due to phase oscillations in the magnetised regions, the field position and number of local minima and maxima that appear in $I_c(f)$ deviate from a non-magnetic junction with non-periodic behaviour. Furthermore, the magnitudes of $I_c$ at local minima are not always 0 and $I_c$ at local maxima do not decrease inversely with $f$ as expected but can even increase. Once the barrier is fully magnetised ($f > 1/2p$), we recover standard $I_c(f)$ behaviour with periodic minima and peaks in $I_c(f)$ with peak heights decreasing inversely with $f$.

The parameters $q$, $p$ and $Q$, influence $I_c(f)$ in different ways. For $Q$ close to 1 the $j_c$ in the magnetized and demagnetized regions closely match, but in the magnetized regions the superconducting phase oscillates faster. In the magnetized regions $I_c$ quadratically decreases with $f$ for small $H$ (f<<1) and the central peak is rounded, resembling a sinc-type function. For $Q$ far from 1 or negative, $j_c$ differs in the magnetized and demagnetized regions. For small $H$ (f<<1), $I_c$ is mainly determined by the propagation of the magnetised region and, because the demagnetised region shrinks linearly, $I_c$ decreases linearly and the central peak is sharp. The difference in the shape of the central peak for $Q = 1$ and $Q = -1$ is demonstrated on Fig 2.

The other two parameters p and q affect the position of the minima and maxima as illustrated in Fig. 3 which shows the position of the local minima and maxima of $I_c(f)$ for multiple sets of parameters. The main effect of *q* on $I_c(f)$ related to the spacing between minima and maxima. In general, higher values of *q* bring minima and maxima closer to the origin (*H* = 0) since the higher permeability in the junction causes the superconducting phase to oscillate faster with *f* in the magnetised regions. However, for *Q* close to 1, some pairs of minima converge towards each other and the maximum between them disappear.

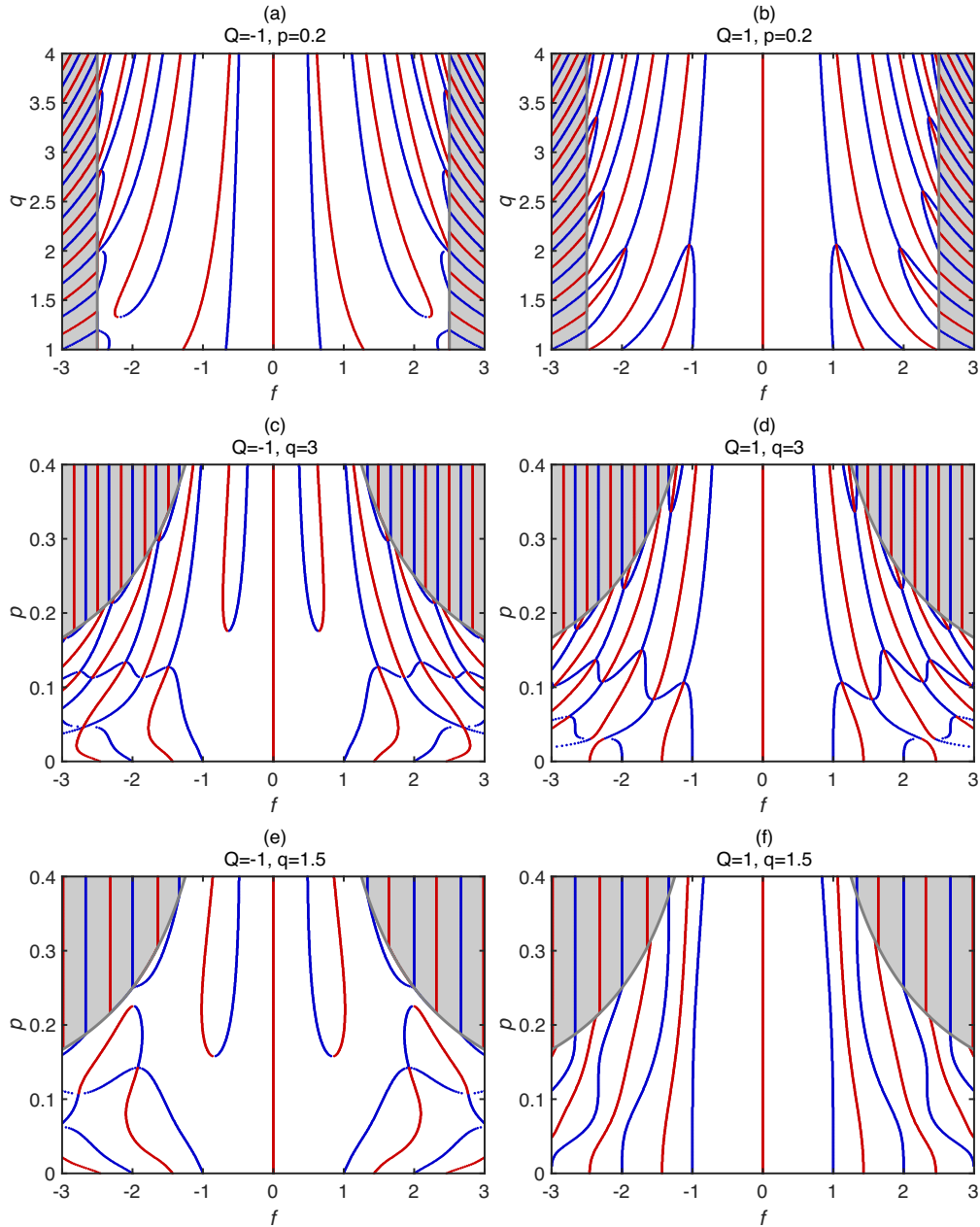

**Fig. 3. The positions of local maxima and minima in *Ic(f)*.** (a) and (b) show the positions of the minima (blue) and the maxima (red) of *Ic(f)* with increasing *q*. (c-f) illustrate the movement of the minima and the maxima as *p* changes for *q* = 3 (c-d), and *q* = 1.5 (e-f). The grey lines indicate the field values where the barrier is magnetised fully. In the fully magnetized regions (shaded grey), a standard *Ic(f)* Fraunhofer behaviour is observed.

The influence of *p* on $I_c(f)$ is most significant for $p < 0.2$. In this range, small changes in *p* affect the shape of $I_c(f)$ significantly: multiple minima in $I_c(f)$ combine and some minima split into two minima forming a maximum (Fig. 3). For *Q* < 0, there is a minimum-maximum pair forming just below $p = 0.2$ (exact value depends on *q* and *Q*). For p > 0.2, the shape of $I_c$(*f*) weakly depends on *p* since the magnetized regions propagate rapidly with *H* and the magnetized regions dominate $I_c$(*f*).

**Conclusions**

We have presented a generalised model to predict the behaviour of $I_C$(*H*) patterns in magnetic Josephson junctions with a non-uniform magnetic susceptibility that peaks at junction edges. An analytical expression for $I_c$(*H*) is derived and key parameters which describe the shape of $I_c$(*H*) are identified: the effective magnetic permeability *q* of the magnetised region; the propagation *p* of the magnetised region into the demagnetized region; and *Q,* the ratio of the local critical current density in the magnetized and demagnetized regions. The calculations can be easily applied to understand the $I_c$(*H*) behaviour magnetically complex Josephson junctions with simultaneous zero and Pi states.

**Acknowledgements**

B. B. Acknowledges funding from the Cambridge Trust and St John's College Cambridge. B. B. and J. W. A. R. acknowledge funding from the Royal Society and the EPSRC through an International Network Grant and Programme Grant (No. EP/P026311/1 and No. EP/ N017242/1). A. B. thanks The Leverhulme Trust for supporting his visiting Professorship in Cambridge University.

**Bibliography**

1. Linder, J. & Robinson, J. W. A. Superconducting spintronics. *Nat. Phys.* **11,** 307–315 (2015).
2. Bergeret, F. S., Volkov, A. F. & Efetov, K. B. Long-range proximity effects in superconductor-ferromagnet structures. *Phys. Rev. Lett.* **86,** 4096–4099 (2001).
3. Bergeret, F. S., Volkov, A. F. & Efetov, K. B. Odd triplet superconductivity and related phenomena in superconductor- ferromagnet structures. *Rev. Mod. Phys.* **77,** 1321–1373 (2005).
4. Buzdin, A. I. Proximity effects in superconductor-ferromagnet heterostructures. *Rev. Mod. Phys.* **77,** 935–976 (2005).
5. Eschrig, M. & Löfwander, T. Triplet supercurrents in clean and disordered half-metallic ferromagnets. *Nat. Phys.* **4,** 138–143 (2008).
6. Eschrig, M. Spin-polarized supercurrents for spintronics. *Phys. Today* **64,** 43–49 (2011).
7. Eschrig, M. Theory of Andreev bound states in S-F-S junctions and S-F proximity devices. *Philos. Trans. R. Soc. A Math. Eng. Sci.* **376,** 20150149 (2018).
8. Ryazanov, V. V. *et al.* Coupling of two superconductors through a ferromagnet: Evidence for a π junction. *Phys. Rev. Lett.* **86,** 2427–2430 (2001).
9. Kontos, T. *et al.* Josephson Junction through a Thin Ferromagnetic Layer: Negative Coupling. *Phys. Rev. Lett.* **89,** 137007 (2002).
10. Blum, Y., Tsukernik, A., Karpovski, M. & Palevski, A. Oscillations of the Superconducting Critical Current in Nb-Cu-Ni-Cu-Nb Junctions. *Phys. Rev. Lett.* **89,** 187004 (2002).
11. Oboznov, V. A., Bol'ginov, V. V., Feofanov, A. K., Ryazanov, V. V. & Buzdin, A. I. Thickness Dependence of the Josephson Ground States of Superconductor-Ferromagnet-Superconductor Junctions. *Phys. Rev. Lett.* **96,** 197003 (2006).
12. Weides, M. *et al.* 0-π Josephson tunnel junctions with ferromagnetic barrier. *Phys. Rev. Lett.* **97,** 247001 (2006).
13. Born, F. *et al.* Multiple 0-π transitions in superconductor/insulator/ferromagnet/ superconductor Josephson tunnel junctions. *Phys. Rev. B - Condens. Matter Mater. Phys.* **74,** 140501 (2006).
14. Piano, S., Robinson, J. W. A., Burnell, G. & Blamire, M. G. 0-π oscillations in nanostructured Nb/Fe/Nb Josephson junctions. *Eur. Phys. J. B* **58,** 123–126 (2007).
15. Robinson, J. W. A., Piano, S., Burnell, G., Bell, C. & Blamire, M. G. Critical current oscillations in strong ferromagnetic π junctions. *Phys. Rev. Lett.* **97,** (2006).
16. Bell, C., Loloee, R., Burnell, G. & Blamire, M. G. Characteristics of strong ferromagnetic Josephson junctions with epitaxial barriers. *Phys. Rev. B - Condens. Matter Mater. Phys.* **71,** 180501 (2005).
17. Robinson, J. W. A., Barber, Z. H. & Blamire, M. G. Strong ferromagnetic Josephson devices with optimized magnetism. *Appl. Phys. Lett.* **95,** 192509 (2009).
18. Bannykh, A. A. *et al.* Josephson tunnel junctions with a strong ferromagnetic interlayer. *Phys. Rev. B* **79,** 054501 (2009).
19. Robinson, J. W. A., Piano, S., Burnell, G., Bell, C. & Blamire, M. G. Zero to π transition

in superconductor-ferromagnet-superconductor junctions. *Phys. Rev. B* **76,** 094522 (2007).

20. Kalcheim, Y. *et al.* Magnetic field dependence of the proximity-induced triplet superconductivity at ferromagnet/superconductor interfaces. *Phys. Rev. B* **89,** 180506 (2014).

21. Kalcheim, Y., Millo, O., Egilmez, M., Robinson, J. W. A. & Blamire, M. G. Evidence for anisotropic triplet superconductor order parameter in half-metallic ferromagnetic La$_{0.7}$Ca$_{0.3}$Mn$_3$O proximity coupled to superconducting Pr$_{1.85}$Ce$_{0.15}$CuO$_4$. *Phys. Rev. B - Condens. Matter Mater. Phys.* **85,** 104504 (2012).

22. Yates, K. A. *et al.* Andreev bound states in superconductor/ferromagnet point contact Andreev reflection spectra. *Phys. Rev. B* **95,** 094516 (2017).

23. Usman, I. T. M. *et al.* Evidence for spin mixing in holmium thin film and crystal samples. *Phys. Rev. B* **83,** 144518 (2011).

24. Kalcheim, Y., Millo, O., Di Bernardo, A., Pal, A. & Robinson, J. W. A. Inverse proximity effect at superconductor-ferromagnet interfaces: Evidence for induced triplet pairing in the superconductor. *Phys. Rev. B* **92,** 060501 (2015).

25. Di Bernardo, A. *et al.* Signature of magnetic-dependent gapless odd frequency states at superconductor/ferromagnet interfaces. *Nat. Commun.* **6,** 8053 (2015).

26. Linder, J. & Robinson, J. W. A. Strong odd-frequency correlations in fully gapped Zeeman-split superconductors. *Sci. Rep.* **5,** 15483 (2015).

27. Blamire, M. G., Smiet, C. B., Banerjee, N. & Robinson, J. W. A. Field modulation of the critical current in magnetic Josephson junctions. *Supercond. Sci. Technol.* **26,** (2013).

28. Anwar, M. S., Veldhorst, M., Brinkman, A. & Aarts, J. Long range supercurrents in ferromagnetic CrO$_2$ using a multilayer contact structure. *Appl. Phys. Lett.* **100,** 052602 (2012).

29. Klose, C. *et al.* Optimization of Spin-Triplet Supercurrent in Ferromagnetic Josephson Junctions. *Phys. Rev. Lett.* **108,** 127002 (2012).

30. Khaire, T. S., Khasawneh, M. A., Pratt, W. P. & Birge, N. O. Observation of Spin-Triplet Superconductivity in Co-Based Josephson Junctions. *Phys. Rev. Lett.* **104,** 137002 (2010).

31. Khasawneh, M. A., Pratt, W. P. & Birge, N. O. Josephson junctions with a synthetic antiferromagnetic interlayer. *Phys. Rev. B* **80,** 020506 (2009).

32. Bakurskiy, S. V. *et al.* Theoretical model of superconducting spintronic SIsFS devices. *Appl. Phys. Lett.* **102,** 192603 (2013).

33. Robinson, J. W. A., Witt, J. D. S. & Blamire, M. G. Controlled injection of spin-triplet supercurrents into a strong ferromagnet. *Science* **329,** 59–61 (2010).

34. Robinson, J. W. A., Chiodi, F., Egilmez, M., Halász, G. B. & Blamire, M. G. Supercurrent enhancement in Bloch domain walls. *Sci. Rep.* **2,** 699 (2012).

35. Houzet, M. & Buzdin, A. I. Long range triplet Josephson effect through a ferromagnetic trilayer. *Phys. Rev. B - Condens. Matter Mater. Phys.* **76,** 060504 (2007).

36. Robinson, J. W. A., Banerjee, N. & Blamire, M. G. Triplet pair correlations and

nonmonotonic supercurrent decay with Cr thickness in Nb/Cr/Fe/Nb Josephson devices. *Phys. Rev. B* **89,** 104505 (2014).

37. Banerjee, N., Robinson, J. W. A. & Blamire, M. G. Reversible control of spin-polarized supercurrents in ferromagnetic Josephson junctions. *Nat. Commun.* **5,** 4771 (2014).
38. Chikazumi, S. *Physics of magnetism*. (R.E. Krieger Pub. Co, 1978).
39. Koehler, W. C., Cable, J. W., Wilkinson, M. K. & Wollan, E. O. Magnetic Structures of Holmium. I. The Virgin State. *Phys. Rev.* **151,** 414–424 (1966).
40. Rhodes, B. L., Legvold, S. & Spedding, F. H. Magnetic Properties of Holmium and Thulium Metals. *Phys. Rev.* **109,** 1547–1550 (1958).
41. Witt, J. D. S. *et al.* Strain dependent selection of spin-slip phases in sputter deposited thin-film epitaxial holmium. *J. Phys. Condens. Matter* **23,** 416006 (2011).
42. Koehler, W. C., Cable, J. W., Child, H. R., Wilkinson, M. K. & Wollan, E. O. Magnetic Structures of Holmium. II. The Magnetization Process. *Phys. Rev.* **158,** 450–461 (1967).
43. Strandburg, D. L., Legvold, S. & Spedding, F. H. Electrical and Magnetic Properties of Holmium Single Crystals. *Phys. Rev.* **127,** 2046–2051 (1962).
44. Gu, Y. *et al.* Magnetic state controllable critical temperature in epitaxial Ho/Nb bilayers. *APL Mater.* **2,** 046103 (2014).
45. Safrata, R. S., Fisher, T. R. & Shelley, E. G. Magnetic Hysteresis of Holmium Metal at 4.2°K. *J. Appl. Phys.* **37,** 4869–4872 (1966).
46. Barone, A. & Paternò, G. *Physics and Applications of the Josephson Effect*. (Wiley and Sons Inc, 1982). doi:10.1002/352760278X
47. Tinkham, M. *Introduction to superconductivity*. (Dover Publications, 1994).
48. Bulaevskii, L. N., Kuzii, V. V & Sobyanin, A. A. Superconducting system with weak coupling to the current in the ground state. *JETP Lett.* **25,** 290 (1977).
49. Buzdin, A. I., Bulaevskil, L. N. & Panyukov, S. V. Critical-current oscillations as a function of the exchange field and thickness of the ferromagnetic metal ( F ) in an S-F-S Josephson junction. *JETP Lett.* **35,** 178 (1982).

**Competing interests**

The authors declare no competing interests.

**Author contributions**

The project was supervised by J. W. R. B. B and A. B performed the calculations and the analysis with support from J. W. R and S. K. The manuscript was written by B. B. and J. W. R. and all authors reviewed it.